\documentclass[a4paper]{article}
\usepackage{latexsym}
\usepackage{graphics}

\begin{document}

\title{Of overlapping Cantor sets and earthquakes:
 Analysis of the discrete Chakrabarti-Stinchcombe model}

\author{Pratip Bhattacharyya\thanks{pratip@cmp.saha.ernet.in}\\ \\
 \small{Theoretical Condensed Matter Physics Division},\\
 \small{Saha Institute of Nuclear Physics},\\
 \small{Sector-1, Block-AF, Bidhannagar, Kolkata 700064, India}}

\date{August 17, 2004}

\maketitle

\begin{abstract}

We report an exact analysis of a discrete form of the Chakrabarti-Stinchcombe
model for earthquakes [Physica A \textbf{270}, 27 (1999)] which considers a pair
of dynamically overlapping finite generations of the Cantor set as a prototype
of geological faults. In this model the $n$-th generation of the Cantor set
shifts on its replica in discrete steps of the length of a line segment in
that generation and periodic boundary conditions are assumed. We determine
the general form of time sequences for the constant magnitude overlaps and
hence obtain the complete time-series of overlaps by the superposition of
these sequences for all overlap magnitudes. From the time-series we derive
the exact frequency distribution of the overlap magnitudes. The corresponding
probability distribution of the logarithm of overlap magnitudes for the $n$-th
generation is found to assume the form of the binomial distribution for $n$
Bernoulli trials with probability $1/3$ for the success of each trial. For
an arbitrary pair of consecutive overlaps in the time-series where the
magnitude of the earlier overlap is known, we find that the magnitude of the
later overlap can be determined with a definite probability; the conditional
probability for each possible magnitude of the later overlap follows the
binomial distribution for $k$ Bernoulli trials with probability $1/2$ for the
success of each trial and the number $k$ is determined by the magnitude of the
earlier overlap. Though this model does not produce the Gutenberg-Richter
law for earthquakes, our results indicate that the fractal structure of faults
admits a probabilistic prediction of earthquake magnitudes.

\vspace{0.5cm}


\vspace{0.5cm}


\end{abstract}

\newpage

\section{Introduction}

\indent Earthquakes are outcomes of fault dynamics in the lithosphere. A
geological fault is comprised of two rock surfaces in contact, created by a
fracture in the rock layers. The two sides of the fault are in slow relative
motion which causes the surfaces to slide. However, owing to friction the
surfaces tend to stick and stress develops in the regions of contact. When
the accumulated stress exceeds the resistance due to friction, the fault
surfaces slip. The potential energy of the strain is thereby released,
causing an earthquake. The slip is eventually stopped by friction and stress
development resumes. Strain continues to develop till the fault surfaces slip
again. This intermittent stick-slip process is the essential feature of fault
dynamics. The overall distribution of earthquakes, including main shocks,
foreshocks and aftershocks, is given by the Gutenberg-Richter
law~\cite{Gutenberg1944, Gutenberg1954}:
\begin{equation}
\log_{10} \mathrm{Nr} ({\cal M}>M) = a - b \: M
\label{eq:Gutenberg-Richter-law}
\end{equation}

\noindent where, $\mathrm{Nr}(\mathcal{M}>M)$ denotes the number (or, the
frequency) of earthquakes of magnitudes ${\cal M}$ that are greater than a
certain value $M$. The constant $a$ represents the total number of
earthquakes of all magnitudes: $a = \log_{10} \mathrm{Nr} (\mathcal{M}>0)$
and the value of the coefficient $b \approx 1$ is presumed to be universal.
In an alternative form, the Gutenberg-Richter law is expressed as a relation
for the number (or, the frequency) of earthquakes in which the energy
released $\mathcal{E}$ is greater than a certain value $E$:
\begin{equation}
\mathrm{Nr} ({\cal E}>E) \sim  E^{-b/\beta}
\label{eq:G-R-power-law}
\end{equation}

\noindent where $\beta \approx 3/2$ is the coefficient in the
energy-magnitude relation~\cite{Knopoff1996, Knopoff2000}.

\indent One class of models for simulating earthquakes is based on the
collective motion of an assembly of connected elements that are driven slowly,
of which the block-spring model due to Burridge and
Knopoff~\cite{Burridge1967} is the prototype. The Burridge-Knopoff model and
its variants~\cite{Carlson1989-1994, Olami1992} have the stick-slip dynamics
necessary to produce earthquakes. The underlying principle in this class of
models is self-organized criticality~\cite{Bak1987-1988}.

\indent Another class of models for simulating earthquakes is
based on overlapping fractals. These models are motivated by the
observation that a fault surface is a fractal object~\cite{Andrews1980,
Aviles1987, Okubo1987, Scholz1989, Sahimi1993}.
Consequently a fault may be viewed as a pair of overlapping fractals.
Fractional Brownian profiles have been commonly used as models of fault
surfaces~\cite{Brown1985, Sahimi1993, Sahimi1996}.
In that case the dynamics of a fault is represented by one Brownian
profile drifting on another and each intersection of the two profiles
corresponds to an earthquake~\cite{Rubeis1996}. However
the simplest possible model of a fault $-$ from the fractal point of
view $-$ was proposed by Chakrabarti and Stinchcombe~\cite{Chakrabarti1999}.
This model is a schematic representation of a fault by a pair of dynamically
overlapping Cantor sets. It is not realistic but, as a system of overlapping
fractals, it has the essential feature. Since the Cantor set is a fractal
with a simple construction procedure, it allows us to study in detail the
statistics of the overlap of one fractal object on another.

\indent In this paper we study the outcome of discrete dynamics in the
Chakrabarti-Stinchcombe model. Our first aim is to construct the time-series
of overlaps of the $n$-th generation of the regular Cantor set on its replica
with periodic boundary conditions, where one set shifts on the other in
discrete steps. We use a finite generation of the Cantor set since
self-similarity in natural objects extends only over a finite range.
In this model the overlap magnitudes and their logarithms correspond
to the energies released and the magnitudes of earthquakes respectively.
Our second aim is to obtain the statistics of the overlap magnitudes.
From the time-series we derive the exact frequency distribution of the
overlap magnitudes. For the $n$-th generation of the Cantor set the
probability distribution of the logarithm of overlap magnitudes is found
to assume the form of a binomial distribution for $n$ Bernoulli trials
and not an exponential law or a power law of the forms expressed
in Eqs.~(\ref{eq:Gutenberg-Richter-law}) and (\ref{eq:G-R-power-law}).
As the Cantor set is approached through successive generations,
the most probable overlap magnitude is found to approach the cube-root
of the maximum overlap magnitude. Our third aim is to determine whether it
is possible to predict the occurrence of an overlap of a particular magnitude
after a certain interval of time from an overlap that is known to have
occurred. For this purpose we apply the theory of conditional probability.
We consider the simplest case, that of two consecutive overlaps
in the time-series, where the magnitude of the earlier overlap is
preassigned and using this we calculate the probabilities of all
possible magnitudes of the later overlap. Each conditional probability
is found to follow a binomial distribution for a certain number
of Bernoulli trials that is determined by the magnitude of the earlier
overlap. The asymptotic form of the most probable magnitude of the later
overlap is found to be inversely proportional to the square-root of the
magnitude of the earlier overlap. We use the expression for the conditional
probability to study three important cases: where the magnitude of the
later overlap is less than that of the earlier one, where it is the
converse of the previous case and where the magnitude of the two overlaps
are equal. On the basis of the fractal overlap analogy between a pair of
fault surfaces and a pair of overlapping Cantor sets, our results indicate
that, from the knowledge of earthquakes recorded in the past it is
possible to predict, with a probability, the magnitude of earthquakes
at least for a short term in the future.

\section{The model: Two dynamically overlapping Cantor sets}

\begin{figure}[htbp]
\resizebox{!}{!}{\rotatebox{0}{\includegraphics{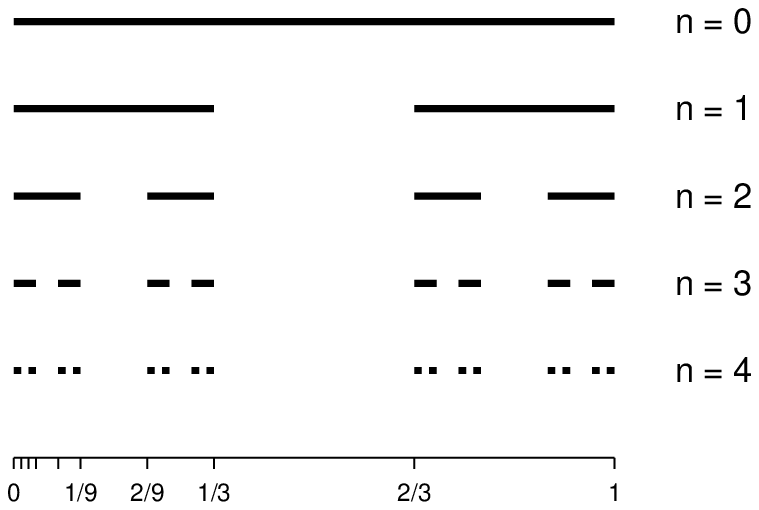}}}
\caption{The \lq initiator\rq ($n=0$) and the first four generations in the
construction of the regular Cantor set. The first generation ($n=1$) provides
the \lq generator\rq. This figure illustrates the process of constructing
successive generations described in the text.}
\label{fig1}
\end{figure}

\indent We consider a discrete form of the model of fault dynamics
proposed by Chakrabarti and Stinchcombe~\cite{Chakrabarti1999}.
The construction of the model requires two identical generations of
the regular Cantor set. The procedure of constructing a regular Cantor
set begins with a line segment of unit length, called the \lq initiator\rq.
This line segment is divided into three equal parts and the middle part
is removed to obtain the first generation; that serves as the
\lq generator\rq~of the Cantor set~\cite{Mandelbrot1982}.
The procedure is repeated for each of the two line segments of
the first generation to obtain the second generation and so on.
Therefore the $n$-th generation of the Cantor set is a finite set
of $2^n$ line segments, each of length $1/3^n$. If this procedure
is repeated an infinite number of times the remainder set of discrete
points is called the regular Cantor set. In the rest of this paper the
term Cantor set will always mean the regular Cantor set that we have just
described. It is an exact self-similar fractal of dimension $\log_3 2$.
The construction of the first few generations is shown in Fig.~\ref{fig1}.
Since it is practically impossible to generate the regular Cantor
set exactly, we use the set of line segments obtained after a finite
number of generations. It is reasonable to work with a finite
generation of the Cantor set since in all naturally occurring
fractals the notion of self-similarity only applies between a lower
and an upper cutoff length~\cite{Mandelbrot1982}.

\indent The model is defined in terms of two dynamically overlapping
fractals which are represented by the $n$-th generation of the Cantor
set and its replica. The sequences of line segments in the two finite sets
of the $n$-th generation are considered as a schematic representation of
the fractal profiles of the two surfaces of the fault. The magnitude of the
overlap of one fault surface on the other is measured in this model as the
number of line segments of one set that overlap exactly on line segments of
the other set and we denote it by $Y_n$ for the $n$-th generation sets.
The motion of the fault surfaces is simulated by shifting one of the sets
relative to the other in the following manner. We assume that initially the
$n$-th generation of the Cantor set and its replica overlap completely, i.e.,
every line segment of one set overlaps exactly on the corresponding line
segment of the other set. Then we shift one of the two finite sets relative
to the other in one particular direction in uniform discrete steps of length
$1/3^n$, that is the length of a line segment in the $n$-th generation. The
length of a step is chosen to be $1/3^n$ in order to ensure that each line
segment of one set either overlaps exactly on one of the other set or does
not overlap at all.  The overlap magnitude $Y_n$ after every step is given by
the number of overlapping segments. Consequently it is convenient to define
time $t$ for this dynamical process as a discrete variable and measure its
value as the number of uni-directional steps by which one set has shifted
from its initial position relative to the other. The time-series of overlaps
is given by the sequence $\{Y_n(t)\}$. Due to the discrete nature of the
shifting process the magnitude of the overlap has discrete values. Further
the structure of the $n$-th generation allows the number of overlapping
segments to be in powers of 2 only: $Y_n = 2^{n-k}$, $k=0, \ldots , n$. We
shall write the magnitude of overlap as $2^{n-k}$ in order to indicate the
generation $n$ to which it belongs. We use periodic boundary conditions which
produces a periodicity in the time-series. In Fig.~\ref{fig3} we plot the
time-series of overlap for the first few generations of the Cantor set.
These plots show that the time-series has the appearance of a self-affine
profile.  

\section {Construction of the time-series}

\indent In the following we construct the time-series of overlaps $Y_n(t)$
of the $n$-th generation of the Cantor set on its replica by determining all
of its structural features. We assume that periodic boundary conditions
are assigned to the finite sets. The structural features of the time-series
consist of a periodicity, a symmetry and a set of distinct temporal sequences
of overlap magnitudes. The most important are the sequences of constant
magnitude overlaps. The complete time-series is derived by superposing the
constant overlap sequences of all possible overlap magnitudes.

\indent (i) Since the initial condition is chosen as the complete overlap
of the two sets, the initial overlap is maximum and its magnitude is given by:
\begin{equation}
Y_n(0) = \max Y_n = 2^n.
\label{eq:initial-overlap}
\end{equation}

\indent (ii) Since the length of each line segment in the $n$-th
generation of the Cantor set is $1/3^n$ and the boundary conditions
are periodic, the time-series of the overlap for the generation
$n$ repeats itself after every $3^n$ time-steps:
\begin{equation}
Y_n(t) = Y_n \left ( t+3^n \right ).
\label{eq:periodicity}
\end{equation}

\noindent Subsequently whenever we shall refer to the variable $t$
and its values it will mean
\begin{equation}
t \equiv t \: \mathrm{mod} \: 3^n,
\label{eq:within-a-period}
\end{equation}

\noindent such that we shall study the time-series within the period
$0 \leq t < 3^n$.

\indent (iii) Since the structure of all generations of the Cantor set
is symmetric about its center and the boundary conditions are periodic,
the time-series of the overlap in the period $0 \leq t < 3^n$
(i.e., in all periods of the kind $k 3^n \leq t < (k+1) 3^n$,
$k = 0, 1, 2, \ldots $) has a symmetric form:
\begin{equation}
Y_n(t) = Y_n \left ( 3^n-t \right ),
 \hspace{0.5cm} 0 \leq t < 3^n.
\label{eq:overlap-symmetry}
\end{equation}

\noindent Therefore, for every sequence of overlaps occurring at time-steps
$\{t\}$ within a period of the time-series, we have a complementary sequence
of overlaps occurring at the time-steps $\{3^n-t\}$ within the same period.

\indent (iv) After the first $3^{n-1}$ time-steps of shifting one
$n$-th generation set relative to the other, the overlapping region
is similar to the generation $n-1$ of the Cantor set. Consequently
the next $3^{n-1}$ time-steps forms a period of the time-series of
overlap of two sets of generation $n-1$. Within this period of $3^{n-1}$
time-steps, after the first $3^{n-2}$ time-steps the overlapping region
is similar to the generation $n-2$ of the Cantor set and therefore
the next $3^{n-2}$ time-steps form a period of the time-series of
overlap of two sets of generation $n-2$.
This process occurs recursively and therefore a period of the
time-series of the generation-$n$ overlap is a nested structure of
the periods of the time-series of the overlaps of all the preceding
generations, for example, after $3^{n-1}+3^{n-2}+\cdots+3^{n-k}$
time-steps the overlapping region is similar to generation $n-k$ of the
Cantor set, for which, following Eq.(\ref{eq:initial-overlap}),
the magnitude of overlap is given by:
\begin{equation}
Y_n \left ( \sum_{r=1}^k 3^{n-r} \right ) = 2^{n-k};
 \hspace{0.5cm} k=1, \ldots ,n.
\label{eq:recursive-overlap}
\end{equation}

\begin{figure}[htbp]
\resizebox{!}{!}{\rotatebox{0}{\includegraphics{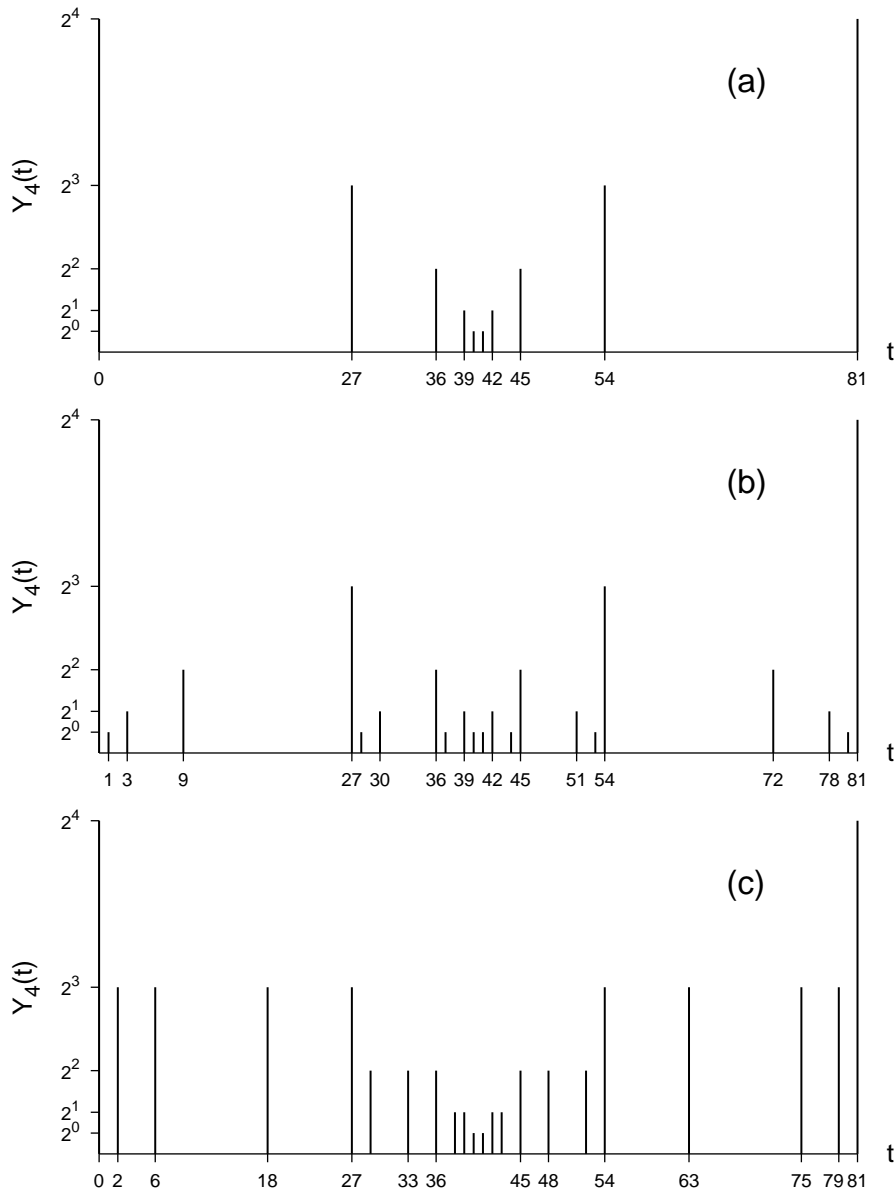}}}
\caption{Three important features of the time-series are illustrated for the
first four generations $(n=1,2,3,4)$ of the Cantor set. The features within
a period of each of the first three generations appear successively nested
within a period of the fourth. (a) The skeleton of the time-series. It is the
nested structure of the maximum overlaps for all the generations.
(b) A sequence of overlaps whose magnitudes are in geometric progression
with common ratio $2$ which occur after the time-steps whose values are in
geometric progression with common ratio 3. (c) The sequence of overlaps of
a constant magnitude that is half the maximum, shown for a period for all
four generations.}
\label{fig2}
\end{figure}

\begin{figure}[htbp]
\resizebox{!}{!}{\rotatebox{0}{\includegraphics{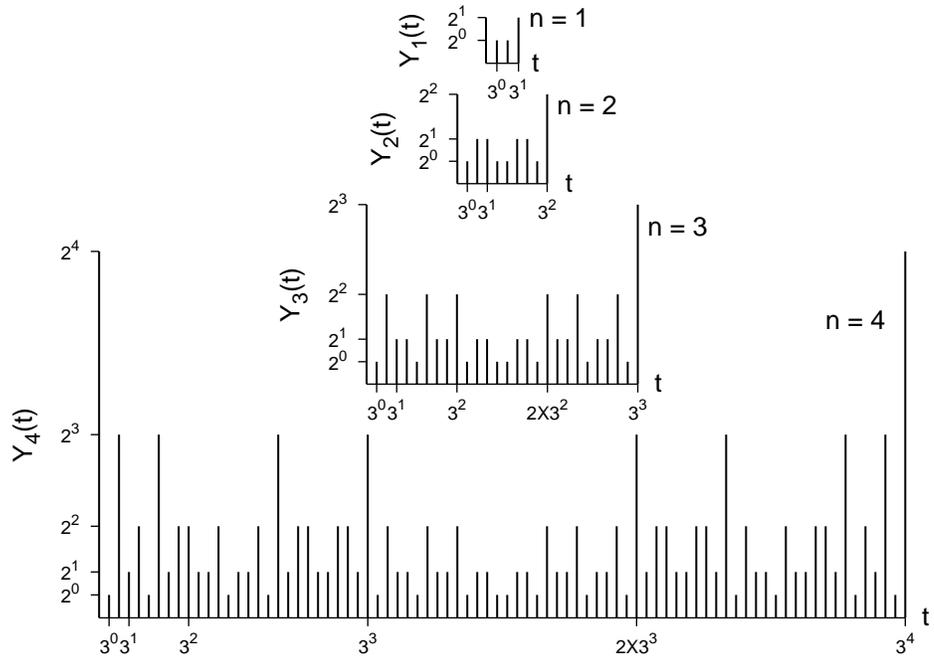}}}
\caption{A period of the time-series of overlap magnitudes for the $n$-th
generation of the Cantor set overlapping on its replica, for the first four
generations $(n = 1, 2, 3 \ {\rm and} \ 4)$ according to
Eqs.~(\ref{eq:initial-overlap}), (\ref{eq:periodicity}),
(\ref{eq:overlap-symmetry}) and (\ref{eq:constant-overlap-2^n-k}). It shows
that a period of the time-series for each generation is nested within a
period of the time-series for the next generation. Owing to this recurrence,
the sequence of overlap magnitudes in a period of the time-series for the
Cantor set $(n=\infty)$ forms a self-affine profile.}
\label{fig3}
\end{figure}

\noindent The sequence of overlaps generated by
Eq.~(\ref{eq:recursive-overlap}) and the complementary sequence obtained
by the symmetry property of Eq.~(\ref{eq:overlap-symmetry}), along with
the initial condition (Eq.~\ref{eq:initial-overlap}), forms
the skeleton of the entire time-series (illustrated in part (a) of
Fig.~\ref{fig2}). The self-affine profile of the time-series observed
in Fig.~\ref{fig3} is due to this property of the overlap. However
it does not provide any more detail of the time-series. 
  
\indent (v) If the generation-$n$ of a Cantor set is shifted by a
single time-step relative to another, the resulting overlap is just
a unit line segment -- owing to the periodic boundary conditions
the last line segment of the former set overlaps on the first line
segment of the latter set. This overlapping segment is similar
to the initiator of the Cantor set (i.e., generation zero) with respect
to the ratio $1/3^n$. Similarly, after a shift by three time-steps,
the last two line segments of the former set overlaps on first two
line segments of the latter set and the overlapping region is similar
to the first generation of the Cantor set with respect to the ratio
$1/3^{n-1}$. In general, owing to the periodic boundary conditions,
a shift of $3^k$ time-steps ($0 \leq k \leq n$) from
the initial position results in the overlapping of the last $2^k$
line segments of one set on the first $2^k$ segments of
the other set and the overlapping region is similar to the
generation-$k$ of the Cantor set with respect to the ratio $1/3^{n-k}$.
Therefore, beginning with a unit overlap resulting after a unit time-step,
a sequence of shifts (time-steps) that are in geometric progression with
$\mathrm{common \ ratio} = 3$ produce a sequence of overlaps with magnitudes
that are in geometric progression with $\mathrm{common \ ratio} =  2$:
\begin{equation}
Y_n \left ( 3^k \right ) = 2^k; \hspace{0.5cm} k=0, \ldots ,n.
\label{eq:geometric-overlap1}
\end{equation}

\noindent This structural feature is illustrated in part (b) of
Fig.~\ref{fig2}. The same geometric progression of overlap magnitudes
also occur for a geometric progression of time-steps beginning with a unit
overlap after the fifth time-step:
\begin{equation}
Y_n \left ( 5 \times 3^k \right ) = 2^k; \hspace{0.5cm} k=0, \ldots ,n.
\label{eq:geometric-overlap2}
\end{equation}

\noindent For each of the sequences given by
Eqs.~(\ref{eq:geometric-overlap1}) and (\ref{eq:geometric-overlap2})
the complementary sequence is obtained by the symmetry property of
Eq.~(\ref{eq:overlap-symmetry}).

\indent (vi) Since the $n$-th generation of the Cantor set contains
$2^n$ line segments that are arranged by the generator in a
self-similar manner, the magnitude of the overlaps, beginning from
the maximum, form a geometric progression of descending powers of 2.
As the magnitude of the maximum overlap is $2^n$ (observed at $t=0$),
the magnitude of the next largest overlap is $2^{n-1}$, i.e.,
a half of the maximum. A pair of nearest line segments form a doublet
and the generation-$n$ of the Cantor set has $2^{n-1}$ such doublets.
Within a doublet, each of the two line segments are two steps away
from the other. Therefore, if one of the sets is shifted from its initial
position by two time-steps relative to the other, only one of the segments
of every doublet of the former set will overlap on the other segment
of the corresponding doublet of the latter set, thus resulting in an
overlap of magnitude $2^{n-1}$. An overlap of $2^{n-1}$ also occurs
if we consider quartets formed of pairs of nearest doublets and
shift one set from its initial position by $2 \times 3$ time-steps relative
to the other. Similarly, in the case of octets formed of pairs of nearest
quartets, a shift of $2 \times 3^2$ time-steps is required to produce
an overlap of $2^{n-1}$. Considering pairs of blocks of $2^{r_1}$ nearest
segments ($r_1 \leq n-1$), an overlap of magnitude $2^{n-1}$ occurs for
a shift of $2 \times 3^{r_1}$ time-steps:
\begin{equation}
Y_n \left ( 2 \times 3^{r_1} \right ) = 2^{n-1}; \hspace{0.5cm}
 r_1=0, \ldots ,n-1.
\label{eq:constant-overlap-2^n-1}
\end{equation}

\noindent The complementary sequence is obtained as usual by the
symmetry property of Eq.~(\ref{eq:overlap-symmetry}). An illustration of
this constant overlap sequence is given in part (c) of Fig.~\ref{fig2}.

\indent Because of the self-similar structure of the Cantor set the
line of argument used to derive Eq.~(\ref{eq:constant-overlap-2^n-1})
will work recursively for overlaps of all consecutive magnitudes.
For example, the next overlap magnitude is $2^{n-2}$, i.e.,
a quarter of the maximum. For each time-step $t_1$ at which an
overlap of $2^{n-1}$ segments occur, there are two subsequences
of overlaps of $2^{n-2}$ segments that are mutually symmetric with
respect to $t_1$; one of the subsequences precedes $t_1$, the other
succeeds $t_1$. Therefore the sequence of $t$ values at which an overlap
of $2^{n-2}$ segments occurs is determined by the sum of two terms, one
from each of two geometric progressions, one nested within the other:
\begin{equation}
Y_n \left ( 2 \left [ 3^{r_1} \pm 3^{r_2} \right ] \right )
 = 2^{n-2};
\label{eq:constant-overlap-2^n-2}
\end{equation}
\begin{eqnarray*}
r_1 & = & 1, \ldots ,n-1; \\
r_2 & = & 0, \ldots ,r_1-1.
\end{eqnarray*}

\noindent The first term belongs to the geometric progression that
determines the sequence of $t$ values appearing in equation
Eq.~(\ref{eq:constant-overlap-2^n-1}) for the overlaps of $2^{n-1}$
segments, while the second term belongs to a geometric progression
nested within the first. The symmetry property of
Eq.~(\ref{eq:overlap-symmetry}) provides the complementary sequence.

\indent In general the sequence of time-step values at which an overlap
of $2^{n-k}$ segments ($1 \leq k \leq n$) occurs is determined
by the sum of $k$ terms, one from each of $k$ geometric progressions,
nested in succession:
\begin{equation}
Y_n \left ( 2 \left [ 3^{r_1} \pm 3^{r_2} \pm \cdots \pm 3^{r_{k-1}}
 \pm 3^{r_k} \right ]
 \right ) = 2^{n-k};
\label{eq:constant-overlap-2^n-k}
\end{equation}
\begin{eqnarray*}
 & k = 1, \ldots, n & ;
\end{eqnarray*}
\begin{eqnarray*}
r_1 & = & k-1, \ldots ,n-1; \\
r_2 & = & k-2, \ldots ,r_1-1; \\
\vdots & & \vdots \\
r_{k-1} & = & 1, \ldots ,r_{k-2}-1; \\
r_k & = & 0, \ldots ,r_{k-1}-1.
\end{eqnarray*}

\noindent For each value of $k$ in the above equation there is a
complementary sequence due to the symmetry property of
Eq.~(\ref{eq:overlap-symmetry}). Eq.~(\ref{eq:constant-overlap-2^n-k})
along with Eqs.~(\ref{eq:initial-overlap}), (\ref{eq:periodicity})
and (\ref{eq:overlap-symmetry}) determine the entire time-series
of overlaps of the $n$-th generation of the Cantor set on its replica.
Assuming that the initial overlap is given by Eq.~(\ref{eq:initial-overlap}),
the time-series is the superposition of the sequences of constant magnitude
overlaps given by Eq.~(\ref{eq:constant-overlap-2^n-k}) for all possible
overlap magnitudes.

\indent Now we consider the special case of unit overlaps, i.e.,
when the overlap is only on a unit segment. We shall write the
magnitude of a unit overlap for the $n$-th generation of the Cantor
set as $2^{n-n}$ in order to indicate the generation index explicitly.
A unit overlap occurs when $k = n$ in Eq.~(\ref{eq:constant-overlap-2^n-k}).
The sequence of $t$ values at which these occur is given by:
\begin{equation}
Y_n \left ( 2 \left [ 3^{n-1} \pm 3^{n-2} \pm \cdots
 \pm 3^1 \pm 3^0 \right ]
 \right ) = 2^{n-n} = 1.
\label{eq:unit-overlap-sequence}
\end{equation}

\noindent The above equation shows $2^{n-1}$ occurrences of the unit
overlap. The same number of unit overlaps also occur in the
complementary sequence obtained by using Eq.~(\ref{eq:overlap-symmetry}).
Therefore, in a period of the time-series for the $n$-th generation,
there are altogether $2^n$ unit overlaps.

\section{Analysis of the time-series}

\indent The analysis of the complete time-series determined by
Eqs.~(\ref{eq:initial-overlap}), (\ref{eq:periodicity}),
(\ref{eq:overlap-symmetry}) and Eq.~(\ref{eq:constant-overlap-2^n-k})
is carried out in two parts. First we systematically derive the exact
number of overlaps of any magnitude $Y_n$ in a period of the time-series.
Next we apply probability theory. The probability of occurrence of an
overlap of magnitude $Y_n$ after any arbitrary time-step is obtained
directly from its frequency. We then derive the conditional probability
of the occurrence of an overlap of magnitude $Y_n'$ if it is given that
the preceding overlap in the time-series has magnitude $Y_n$.

\subsection{Frequency of overlap magnitudes}

\indent In the following ${\rm Nr}(Y_n)$ denotes the frequency of
overlaps of magnitude $Y_n$, i.e., the number of times an overlap of
$Y_n$ segments occurs in a period of the time-series for $n$-th generation
of the Cantor set.
From the sequences defined by Eqs.~(\ref{eq:initial-overlap}),
(\ref{eq:constant-overlap-2^n-1}) and (\ref{eq:constant-overlap-2^n-2})
and their complementary parts, we get respectively:
\begin{equation}
\mathrm{Nr} \left ( 2^n \right ) = 1,
\label{eq:number-fulloverlap}
\end{equation}
\begin{equation}
\mathrm{Nr} \left ( 2^{n-1} \right ) = 2n
\label{eq:number-halfoverlap}
\end{equation}
\noindent and
\begin{equation}
\mathrm{Nr} \left ( 2^{n-2} \right ) = 2 \sum_{r_1=1}^{n-1} 2 \, r_1
 = 2n(n-1).
\label{eq:number-quarteroverlap}
\end{equation}

\noindent Similarly we can calculate the frequency of overlaps
of all magnitudes. For example,
\begin{eqnarray}
\mathrm{Nr} \left ( 2^{n-3} \right ) & = & 2 \sum_{r_1=2}^{n-1}
 2 \sum_{r_2=1}^{r_1-1} 2 \, r_2 \nonumber \\
 & = & {4 \over 3}n(n-1)(n-2),
\label{eq:number-1/8-overlap}
\end{eqnarray}
\begin{eqnarray}
\mathrm{Nr} \left ( 2^{n-4} \right ) & = & 2 \sum_{r_1=3}^{n-1}
 2 \sum_{r_2=2}^{r_1-1} 2 \sum_{r_3=1}^{r_2-1} 2 \, r_3 \nonumber \\
 & = & {2 \over 3}n(n-1)(n-2)(n-3),
\label{eq:number-1/16-overlap}
\end{eqnarray}
\begin{eqnarray}
\mathrm{Nr} \left ( 2^{n-5} \right ) & = & 2 \sum_{r_1=4}^{n-1}
 2 \sum_{r_2=3}^{r_1-1} 2 \sum_{r_3=2}^{r_2-1} 2 \sum_{r_4=1}^{r_3-1}
 2 \, r_4 \nonumber \\
 & = & {4 \over 15}n(n-1)(n-2)(n-3)(n-4).
\label{eq:number-1/32-overlap}
\end{eqnarray}

\noindent In general, from Eq.~(\ref{eq:constant-overlap-2^n-k})
the frequency of overlaps of magnitude $Y_n=2^{n-k}$ is given by:
\begin{eqnarray}
\mathrm{Nr} \left ( 2^{n-k} \right )
 & = & 2 \sum_{r_1=k-1}^{n-1} 2 \sum_{r_2=k-2}^{r_1-1} \cdots \:
       2 \sum_{r_{k-1}=1}^{r_{k-2}-1} 2 \, r_{k-1} \nonumber \\
 & = & C_k \: n(n-1)(n-2) \: \cdots \: (n-k+1) \nonumber \\
 & = & C_k \: {n! \over (n-k)!}.
\label{eq:number-1/2^k-overlap}
\end{eqnarray}

\noindent The value of the constant $C_k$ is determined from the case
of unit overlaps in the following way. In the above equation we keep
the index $k$ constant and choose the generation index $n=k$.
As a result we get the frequency of unit overlaps for the $k$-th
generation:
\begin{equation}
\mathrm{Nr} \left ( 2^{k-k} \right ) = C_k \: k!.
\label{eq:unit-overlap-freque}
\end{equation}

\noindent (We have followed the notation explained before
Eq.~(\ref{eq:unit-overlap-sequence}) that the magnitude of the
unit overlap for the $k$-th generation of the Cantor set is written
as $2^{k-k}$ in order to indicate the generation index.)
On the other hand, we get the frequency of the unit overlap
for the $k$-th generation from the sequence defined in
Eq.~(\ref{eq:unit-overlap-sequence}) by replacing the index $n$ by $k$:
\begin{equation}
\mathrm{Nr} \left ( 2^{k-k} \right ) = 2^k.
\label{eq:unit-overlap-freq}
\end{equation}

\noindent Comparing Eqs.~(\ref{eq:unit-overlap-freque}) and
(\ref{eq:unit-overlap-freq}) we get the value of the constant:
\begin{equation}
C_k = {2^k \over k!}.
\end{equation}

\noindent Now we have the complete formula for the frequency of
overlaps of magnitude $Y_n=2^{n-k}$ in the time-series for the $n$-th
generation of the Cantor set:
\begin{eqnarray}
\mathrm{Nr} \left ( 2^{n-k} \right ) & = & {2^k \over k!} \: {n! \over (n-k)!}
 \nonumber \\
 & = & 2^k \left ( \begin{array}{c}
                            n\\
                            k
                   \end{array} \right ),
 \hspace{0.5cm} k = 0, \ldots, n
\label{eq:number-overlap-2^n-k}
\end{eqnarray}

\noindent where $\left ( \begin{array}{c} n\\ k \end{array} \right )$
denotes the binomial coefficient. The frequency distribution
$\mathrm{Nr}(Y_n)$ obtained in Eq.~(\ref{eq:number-overlap-2^n-k}) above
for the overlap magnitudes $Y_n=2^{n-k}$, $k = 0, \ldots , n$, is exact.
It is our central result and all subsequent results in this paper will be
derived from it. It differs from the earlier claims
~\cite{Chakrabarti1999, Pradhan2003} that the frequency distribution
follows a power law; possible reason for this difference will be mentioned
in the discussion. 

\begin{figure}[htbp]
\resizebox{!}{!}{\rotatebox{0}{\includegraphics{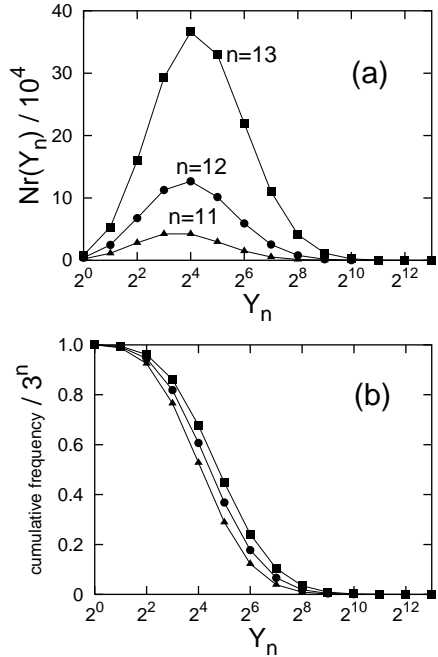}}}
\caption{(a) The frequency distribution of the overlap magnitudes $Y_n
=2^{n-k}$, $k = 0, \ldots , n$ for three consecutive generations $n = 11, 12,
\ \mathrm{and} \ 13$ according to Eq.~(\ref{eq:number-overlap-2^n-k}). The
points have been joined by lines only to make out the asymmetric Gaussian
appearance of the distributions. (b) The cumulative distribution of overlap
magnitudes for the three consecutive generations shown in part (a),
calculated by using Eq.~(\ref{eq:cumulative}). The cumulative distribution
for each generation has been scaled by the period of the corresponding
time-series.}
\label{fig4}
\end{figure}

\indent Using the binomial theorem, we have for the entire period:
\begin{equation}
\sum_{k=0}^n \mathrm{Nr} \left ( 2^{n-k} \right)
 = \sum_{k=0}^n 2^k \left ( \begin{array}{c}
                                      n\\
                                      k
                            \end{array} \right )
 = 3^n,
\label{eq:total-overlap-number}
\end{equation}

\noindent since there are $3^n$ time-steps in a period of the time-series.
From Eq.~(\ref{eq:number-overlap-2^n-k}) we can
calculate the cumulative frequency of overlaps with magnitudes that
are greater than a preassigned value, say, $2^{n-k}$, by a partial sum
of binomial terms:
\begin{equation}
\mathrm{Nr} \left ( Y_n \geq 2^{n-k} \right )
 = \sum_{r=0}^k \mathrm{Nr} \left ( 2^{n-r} \right)
 = \sum_{r=0}^k 2^r \left ( \begin{array}{c}
                                    n\\
                                    r
                          \end{array} \right ),
\label{eq:cumulative}
\end{equation}

\noindent which does not have a closed form. Fig.~\ref{fig4} shows the
frequency distribution and cumulative distribution of overlap magnitudes for
three consecutive generations.  For large overlap magnitudes, which is
realized for large values of $n$ and $k \ll n$,
Eq.~(\ref{eq:number-overlap-2^n-k}) reduces to the following asymptotic form:
\begin{equation}
\mathrm{Nr} \left ( 2^{n-k} \right ) \sim {2^k \: n^k \over k!}.
\label{eq:asymptotic-number-overlap-2^n-k}
\end{equation}

\noindent This is evident from the expressions for $k=0, \ldots ,5$
in Eqs.~(\ref{eq:number-fulloverlap})--(\ref{eq:number-1/32-overlap}).
Using the notations $Y_n = 2^{n-k}$ and $\max Y_n = 2^n$,
Eq.~(\ref{eq:asymptotic-number-overlap-2^n-k}) for the $n$-th generation
may be written as:
\begin{equation}
\mathrm{Nr} \left ( Y_n \right ) \sim \left ( {\max Y_n \over Y_n} \right )
 ^{1 + \log_2 n} \: {1 \over \left ( \log_2 \left [ (\max Y_n) / Y_n \right ]
 \right ) !}
\label{eq:alternative-asymptotic-number-overlap-2^n-k}
\end{equation}

\noindent which shows that the factor $1/Y_n$, which was the one obtained
in~\cite{Chakrabarti1999}, is overwhelmed by the factor $1/Y_n^{\log_2 n}$
in the expression for $\mathrm{Nr} \left ( Y_n \right )$ for large $n$ and
small $k$.

\subsection{Prediction probability of overlap magnitudes}

\indent We now analyze the time-series of overlap from the point
of view of probability theory. The treatment by probability theory
is necessary to determine whether the occurrence of an overlap of
a certain magnitude can be predicted from the knowledge of the
magnitude of an overlap that has already occurred. We continue to
consider the case of the $n$-th generation of the Cantor set
overlapping on its replica.
Let $\Pr(Y_n)$ denote the probability that after any arbitrary
time-step $t$ we get an overlap of $Y_n$ segments. For the general case,
$Y_n=2^{n-k}$, $k=0, \ldots ,n$, it is given by:
\begin{eqnarray}
\Pr \left ( 2^{n-k} \right ) & = & {\mathrm{Nr} \left ( 2^{n-k} \right )
 \over \sum_{k=0}^n \mathrm{Nr} \left ( 2^{n-k} \right)} \nonumber \\
 & = & {2^k \over 3^n} \left ( \begin{array}{c}
                                        n\\
                                        k
                               \end{array} \right ) \nonumber \\
 & = & \left ( \begin{array}{c} n\\ n-k \end{array} \right )
 \left ( {1 \over 3} \right )^{n-k} \left ( {2 \over 3} \right )^k.
\label{eq:overlap-prob}
\end{eqnarray}

\noindent The final expression in the above equation has the form
of the binomial distribution for $n$ Bernoulli trials~\cite{Feller1968}
with probability $1/3$ for the success of each trial and therefore
$\Pr \left ( Y_n \right )$ stands for the probability of the case
where the number of successes is given by $\log_2 \! Y_n$.
Since $\Pr \left ( 2^{n-k} \right )$ is maximum for
$k = \left \lfloor 2(n+1)/3 \right \rfloor$ (and also for $k=2(n+1)/3-1$
when $n+1$ is a multiple of 3), the most probable overlap magnitude
in general is given by:
\begin{equation}
\widehat{Y}_n = 2^{n- \left \lfloor 2(n+1)/3 \right \rfloor}
\label{eq:most-prob-overlap}
\end{equation}

\noindent where the floor function $\lfloor x \rfloor$ is defined as
the greatest integer less than or equal to $x$~\cite{Graham1994}.
Since $\max Y_n = 2^n$, we get the following asymptotic relation
for large values of the generation index $n$:
\begin{equation}
\widehat{Y}_n \sim \left ( \max Y_n \right ) ^{1/3}.
\label{eq:most-prob-overlap-asymp}
\end{equation}

\indent Next we consider the conditional probability that an overlap
of magnitude $Y_n'$ occurs after the time-step $t+1$ if it is known
that an overlap of magnitude $Y_n$ has occurred after the previous
time-step $t$, for any arbitrary $t$. The conditional probability
is given by:
\begin{equation}
\Pr \left ( Y_n' \left \vert \right . Y_n \right )
 = {\mathrm{Nr} \left ( Y_n, Y_n' \right ) \over
   \mathrm{Nr} \left ( Y_n \right ) }
\label{eq:pair-prob}
\end{equation}

\noindent where $\mathrm{Nr} \left ( Y_n, Y_n' \right )$ is the number of
ordered pairs of consecutive overlaps $\left ( Y_n, Y_n' \right )$
occurring in a period of the time-series for the $n$-th generation.
It follows from Eq.~(\ref{eq:constant-overlap-2^n-k}) that overlaps of
magnitude $2^{n-k}$ in the time-series are immediately succeeded
(i.e., after the next time-step) by overlaps of magnitude $2^r$,
$0 \leq r \leq k$ only and never by an overlap of magnitude greater
than $2^k$. For $Y_n=2^{n-k}$ and $Y_n'=2^r, \; r=0, \ldots , k$ we have:
\begin{equation}
\mathrm{Nr} \left ( 2^{n-k}, 2^r \right )
 = \left ( \begin{array}{c} n\\ k \end{array}  \right )
   \left ( \begin{array}{c} k\\ r \end{array}  \right ).
\label{eq:pair-number}
\end{equation}

\noindent Therefore, from Eqs.~(\ref{eq:number-overlap-2^n-k}),
(\ref{eq:pair-prob}) and (\ref{eq:pair-number}) we get:
\begin{eqnarray}
\Pr \left ( 2^r \left \vert \right . 2^{n-k}  \right )
 & = & {1 \over 2^k} \left ( \begin{array}{c}
                                      k\\ r \end{array} \right );
 \hspace{0.5cm} \begin{array}{c} k = 0, \ldots , n\\
                                 r = 0, \ldots , k \end{array} \nonumber \\
 & = & \left ( \begin{array}{c} k\\ r \end{array} \right )
 \left ( {1 \over 2} \right )^r \left ( {1 \over 2} \right )^{k-r}.
\label{eq:pair-prob-cantor}
\end{eqnarray}

\noindent Here we find that the conditional probability follows
the binomial distribution for $k$ Bernoulli trials with probability
$1/2$ for the success of each trial.
Eq.~(\ref{eq:pair-prob-cantor}) shows that the expression of
$\Pr (2^r \vert 2^{n-k})$ is independent of the generation index $n$.
Therefore, for fixed $k$ and $r$, it has the same value for all
generations $n$, provided that $0 \leq r \leq k \leq n$.
Since the conditional probability $\Pr (2^r \vert 2^{n-k})$ is
maximum when $r = \lfloor (k+1)/2 \rfloor$ (and also for $r=(k+1)/2-1$
when $k+1$ is a multiple of 2), the most probable
overlap magnitude $\widehat{Y}_n'$ to occur next to an overlap of
magnitude $Y_n = 2^{n-k}$ is given by:
\begin{equation}
\widehat{Y}_n' = 2^{\left \lfloor (k+1)/2 \right \rfloor}.
\label{eq:most-prob-pair}
\end{equation}

\noindent For large values of the generation index $n$ and large $k$,
$k \leq n$, from the above equation we get the following asymptotic
relation:
\begin{equation}
\widehat{Y}_n' \sim \sqrt{\max Y_n \over Y_n}.
\label{eq:most-prob-pair-asymp}
\end{equation}

\indent We consider three applications of Eq.~(\ref{eq:pair-prob-cantor})
assuming in each case that $Y_n=2^{n-k}$ and $Y_n'$ is the overlap
magnitude occurring next to $Y_n$ in the time-series. First, we find that
\begin{eqnarray}
\Pr (Y_n' \leq Y_n \left \vert \right . Y_n)
 & = & \sum_{r=0}^k \Pr \left ( 2^r \left \vert \right .  2^{n-k} \right )
       \nonumber \\
 & = & 1 \hspace{0.5cm} \mathrm{for} \ 0 \leq k \leq {n \over 2}.
\label{eq:prob-less}
\end{eqnarray}

\noindent This implies that, for $k \leq n/2$, an overlap of magnitude
$Y_n=2^{n-k}$ is always followed by an overlap of equal or less magnitude
$Y_n'$. Consequently the case of an overlap of magnitude greater than
that of the preceding one (i.e., $Y_n'>Y_n$) appears only for $k>n/2$.
Second, we have
\begin{eqnarray}
\Pr (Y_n' \geq Y_n \left \vert \right . Y_n)
& = & \sum_{r=n-k}^k \Pr \left ( 2^r \left \vert \right .  2^{n-k} \right )
       \nonumber \\
& = & {1 \over 2^k} \sum_{r=n-k}^k \left ( \begin{array}{c} k \\ r \end{array}
                                   \right ) 
      \hspace{0.5cm} \mathrm{for} \ {n \over 2} \leq k \leq n.
\label{eq:prob-greater}
\end{eqnarray}

\noindent Since the final expression in the above equation involves a
partial sum of binomial coefficients, it does not have a closed form,
except for the trivial case $k=n$ and a few other special cases.
Therefore it must be calculated numerically for specific values
of $n$ and $k$. Third, if the magnitude of overlap after a certain
time-step is $Y_n=2^{n-k}$, the probability that an overlap of the same
magnitude also occurs after the next time-step is given by:
\begin{eqnarray}
\Pr \left ( Y_n'=Y_n \left \vert \right . Y_n \right )
& = & \Pr \left ( 2^{n-k} \left \vert \right . 2^{n-k} \right ) \nonumber \\
& = & \left \{ \begin{array}{c c} 0 \hfill , & 0 \leq k < {n/2}\\
                                  {1 \over 2^k} \left ( \begin{array}{c}
                                                               k\\ n-k
                                                         \end{array} \right ),
 & n/2 \leq k \leq n \end{array} \right . .
\label{eq:prob-equal-cantor}
\end{eqnarray}

\noindent In this way we can derive the probability of various cases
where $Y_n'$ is specifically related to $Y_n$.

\section{Discussion}

\indent In this paper we have reported the construction and analysis
of the time-series of overlaps for the discrete Chakrabarti-Stinchcombe
model where the $n$-th generation of the Cantor set shifts on its
replica, with periodic boundary conditions, in discrete uniform steps.
The $n$-th generation of the Cantor set is effectively a fractal for
lengths between $1$ and $1/3^n$ (or, between $1$ and $3^n$, if the length
of a line segment in the $n$-th generation is taken as the unit) and
therefore it has no characteristic scale within this range; this is also
reflected in the structure of the time-series within a period. However
the frequencies and the corresponding probabilities of the binary logarithm
of overlap magnitudes follow a binomial distribution, which has a
characteristic scale given by the most probable overlap magnitude. The
existence of a most probable value in the frequency distribution creates the
possibility of predicting overlap magnitudes in the time-series.
We have shown the utility of this simple model in predicting an event from
the knowledge of the preceding event. Though it is believed that earthquakes
cannot be predicted~\cite{Geller1997}, the analysis presented in the previous
section indicates that the overlapping fractal structure of faults ought to
admit probabilistic predictions of earthquake magnitudes. It is left to be
explored how the probability of predicting an overlap magnitude in this model
increases with the width of the preceding interval of time-steps within which
all the overlap magnitudes are known.

\indent The Chakrabarti-Stinchcombe model, proposed for studying the
mechanism of earthquakes, is by far the simplest model containing the
rudiments of the overlapping structure of a geological fault.
The authors of the original model~\cite{Chakrabarti1999} analysed, using
renormalization group argument, the case of the Cantor set shifting
continuously over its replica with open boundary conditions and reported that
the density of the overlap magnitudes $Y$, in the infinite generation limit
($n \to \infty$), was in the form of a continuous power law:
$\rho(Y) \propto 1/Y$. However the discrete form of the model defined for
finite generations of the Cantor set with periodic boundary conditions
-- the subject of this paper -- has a binomial distribution of the binary
logarithm of overlap magnitudes (Eq.~\ref{eq:number-overlap-2^n-k} or
Eq.~\ref{eq:overlap-prob}). If, instead of periodic boundary conditions,
open boundary conditions are used, the time-series will be of finite duration,
consisting of $3^n$ time-steps; with the initial condition given by the
position of maximum overlap at $t=0$, as stated in
Eq.~(\ref{eq:initial-overlap}), there will be no overlapping line segments of
the two $n$-th generation sets for all odd values of $t$ and the frequency of
each overlap magnitude, except the maximum, will be only half of the value
given by Eq.~(\ref{eq:number-overlap-2^n-k}). For large generation indices
the asymptotic form of the binomial distribution is given by the continuous
normal approximation~\cite{Feller1968}, but it never acquires the form of a
power law. The difference in the results of Ref.~\cite{Chakrabarti1999} and
this paper is apparently due to the difference in the nature of the shifting
process which is continuous in the former and discrete in the latter (with
the size of each discrete step being equal to the length of the line segment
in the finite generation). The authors of Ref.~\cite{Chakrabarti1999} also
report that their result is valid for all types of Cantor sets and in
fractals of higher dimensions, e.g., in the case of two overlapping
Sierpinsky carpets. Though the discrete forms of these cases are yet to be
studied, it has been reported in~\cite{Pradhan2003} that for two spanning
clusters at the percolation threshold on a square lattice -- where each
cluster is a random fractal embedded in two dimensions --  the overlap
magnitudes follow a normal distribution.

\indent Finally we consider a variant of the Chakrabarti-Stinchcombe model
defined as the $n$-th generation of the Cantor set overlapping on its
complement in the unit line segment; the complement of the $n$-th generation
set is obtained by replacing each line segment (of length $1/3^n$) in the
latter by an empty segment and each empty segment by a line segment. The
overlap magnitudes in this variant model are given by $Y_n = 2^n - 2^k$,
$k=0, \ldots ,n$. The time-series of overlaps can be directly derived from
Eq.~(\ref{eq:constant-overlap-2^n-k}) by replacing each overlap magnitude
$2^{n-k}$ in the time-series of the original model with an overlap magnitude
$2^n-2^{n-k}$. Consequently the frequency distribution of overlap magnitudes
is given by $\mathrm{Nr} \left ( 2^n - 2^k \right ) = 2^{n-k} \left ( 
\begin{array}{c} n\\ k \end{array} \right )$. The variant appears to be more
realistic than the original model since each of the two parts of a fractured
rock is complementary to the other. However, in both cases the frequency of
overlap magnitudes are found to follow binomial distributions.

\section*{Acknowledgment}
I am grateful to Bikas K. Chakrabarti and Robin B. Stinchcombe for their
comments on the manuscript.

\vspace{2.0cm}

\end{document}